\documentclass[11pt,a4paper]{article}
\pdfoutput=1

\usepackage{jheppub}
\usepackage[dvipsnames]{xcolor}
\usepackage{mathrsfs}
\usepackage{amsmath,amsthm,amssymb,slashed}
\usepackage{color}
\usepackage{multirow}
\usepackage[normalem]{ulem}
\usepackage{afterpage}

\newcommand{\be}{\begin{equation}}
\newcommand{\ee}{\end{equation}}
\newcommand{\bea}{\begin{eqnarray}}
\newcommand{\eea}{\end{eqnarray}}
\newcommand{\ba}{\begin{array}}
\newcommand{\ea}{\end{array}}

\newcommand{\muew}{\mu_{w}}

\newcommand{\ie}{\emph{i.e., }}
\newcommand{\eg}{\emph{e.g., }}

\allowdisplaybreaks

\title{New bounds on axion-like particles from MicroBooNE 
}

\author[a]{Pilar Coloma,}
\affiliation[a]{Instituto de F\'isica Te\'orica UAM-CSIC, Calle de
  Nicol\'as Cabrera 13--15, Universidad Aut\'onoma de Madrid,
  Cantoblanco, E-28049 Madrid, Spain}
\emailAdd{pilar.coloma@ift.csic.es}

\author[b]{Pilar Hern\'andez,}
\affiliation[b]{Instituto de F\'isica Corpuscular, Universidad de Valencia \& CSIC, 
Edificio Institutos de Investigaci\'on, Calle Catedr\'atico Jos\'e Beltr\'an 2, 46980 Paterna, Spain}
\emailAdd{m.pilar.hernandez@uv.es}

\author[b]{Salvador Urrea}
\emailAdd{salvador.urrea@ific.uv.es}

\abstract{
Neutrino experiments lie at the edge of the intensity frontier and therefore can be exploited to search for new light particles weakly coupled to the visible sector. In this work we derive new constraints on axion-like particles (ALPs) using data from the MicroBooNE experiment, from a search for $e^+ e^-$ pairs pointing in the direction of the NuMI absorber. In particular, we consider the addition of higher-dimensional effective operators coupling the ALP to the electroweak gauge bosons. These would induce $K\to \pi a$ from kaon decay at rest in the NuMI absorber, as well as ALP decays into pairs of leptons or photons. We discuss in detail and compare various results obtained for the decay width $K\rightarrow \pi a$ in previous literature. For the operator involving the Higgs, MicroBooNE already sets competitive bounds (comparable to those of NA62) for ALP masses between 100 and 200 MeV. We also compute the expected sensitivities from the full NuMI dataset recorded at MicroBooNE. Our results show that a search for a $a\to\gamma\gamma$ signal may be able to improve over current constraints from beam-dump experiments on the operator involving the ALP coupling to the $W$.
}

\preprint{FTUV-22-0121.2990, IFT-UAM/CSIC-22-4, IFIC/22-03}


\begin{document}

\maketitle

\section{Introduction}

Light pseudoscalars may arise as pseudo-Nambu-Goldstone bosons of a spontaneous global symmetry breaking, and are therefore ubiquitous in extensions of the Standard Model (SM). These particles as often referred to as axion-like particles (ALPs) since the best motivated example is the axion \cite{Weinberg:1977ma,Wilczek:1977pj},  in models that  solve the strong CP problem and could explain dark matter~\cite{Preskill:1982cy,Abbott:1982af,Dine:1982ah}. If the scale of symmetry breaking, $f_a$, is much higher than the electroweak (EW) scale, these particles might be the only accessible BSM relics of the high-energy theory. 

In recent years, model-independent constraints on these particles have been systematically studied  using an Effective Field Theory (EFT) framework~\cite{Georgi:1986df,Izaguirre:2016dfi,Gavela:2019wzg,Bauer:2020jbp,Chala:2020wvs,Bauer:2021wjo,Bauer:2021mvw,Bonilla:2021ufe,Guerrera:2021yss}, that includes all the possible couplings of the ALPs allowed by the SM symmetries and the shift symmetry. At the renormalizable level, only a soft-breaking ALP mass term is included,  while the leading non-trivial interactions are of dimension five, and  therefore suppressed  as $f^{-1}_a$. As a result, ALPs fall in the category of \emph{feebly interacting particles}. Although the Peccei-Quinn solutions to the strong CP problem \cite{Peccei:1977hh} imply a very restrictive relation between the axion mass\footnote{The physical mass depends on the tree-level mass arising in some UV extensions, and generically on the non-perturbative QCD contribution to the axion potential. } and $f_a$, fixed by QCD dynamics, alternative scenarios have been considered where this relation may be significantly modified, increasing the target parameter space accordingly (see for example Ref.~\cite{DiLuzio:2020wdo} for a recent review). 

The experimental search for axions and ALPs has intensified enormously in recent years. This has been driven in part by the operation of new axion search experiments, but also by the identification of new and complementary opportunities offered by experiments in the intensity frontier at accelerator facilities, both in beam-dumps and colliders. Stringent limits have been recently obtained from $B$-meson decays~\cite{Belle-II:2020jti,LHCb:2015nkv,LHCb:2016awg}, from kaon decays~\cite{BNL-E949:2009dza,NA62:2021zjw} and from neutrino experiments~\cite{Essig:2010gu}. Very important constraints from past beam-dump experiments~\cite{CHARM:1983ayi,Bjorken:1988as} have also been recasted in the context of more general ALP searches~\cite{Dolan:2017osp}. See \eg Ref.~\cite{Goudzovski:2022vbt} for a recent review of ALP bounds in the region around the kaon mass, or Ref.~\cite{MartinCamalich:2020dfe} for a recent review of flavor constraints on the QCD axion. Furthermore many future experiments are being proposed that can improve the limits further (for a recent review see Ref.~\cite{Agrawal:2021dbo}). 

In this paper we derive a new constraint on ALPs from recent data by the MicroBooNE experiment \cite{MicroBooNE:2021usw}. In neutrino experiments using a conventional beam (such as MicroBooNE), most of the mesons produced in the proton collisions in the target are focused by the magnetic horns and decay within the decay pipe. However, a fraction of the protons (which can be as large as 10-15\%) does not get stopped by the target and ends up hitting the absorber at the end of the decay pipe. The kaons produced in such collisions, after losing energy as they interact with the medium, eventually decay at rest within the absorber. Although the fraction of protons which reach the absorber is much smaller than the ones hitting the target, the detector acceptance will be much larger simply because of its proximity. This fact was precisely used for the analysis performed in Ref.~\cite{MicroBooNE:2021usw}, where the MicroBooNE collaboration searched for monoenergetic scalars coming from the direction of the NuMI hadron absorber, located at a distance of only 100 m from the detector. 

In this work, we concentrate on the ALP couplings to EW gauge bosons and on the most relevant ALP production channel, $K\rightarrow \pi a$. The theoretical prediction of this process has been recently revisited in Refs.~\cite{Bauer:2021wjo,Bauer:2021mvw}, including running effects and treating the long-distance dynamics in the context of the chiral effective theory, as originally proposed in \cite{Georgi:1986df}. Alternative estimates for this channel have been used before in Refs.~\cite{Izaguirre:2016dfi,Gavela:2019wzg,Guerrera:2021yss}. We compare the two approaches in detail, and use the results of Ref.~\cite{Bauer:2021wjo} to derive our bounds. We then provide a recasting of the results of the MicroBooNE analysis in Ref.~\cite{MicroBooNE:2021usw} and re-derive a competitive bound on the effective EW couplings of ALPs, using the same decay topology as in their analysis. 

The paper is organized as follows. In Sec.~\ref{sec:chipt} we review the ALP EFT as well as the computation of $K\rightarrow \pi a$ when EW couplings are present. Section~\ref{sec:bounds} discusses the most relevant bounds on this scenario. Our results are then presented in Sec.~\ref{sec:results}: first, we derive new bounds using published MicroBooNE data, comparing them to current constraints; next, we compute the projected sensitivity using the full statistics expected, and study the interference effects between different effective operators. Our summary and conclusions are then presented in Sec.~\ref{sec:conclusions}. Appendices~\ref{app:integrals} and~\ref{app:loop} contain technical details relevant for the discussion in Sec.~\ref{sec:chipt}. 

\section{Electroweak Lagrangian with effective ALP interactions}
\label{sec:chipt}

We  consider the following set of effective operators describing ALP interactions with EW bosons at some high scale $\Lambda$ (which we take as  $\Lambda = f_a$)
\begin{equation}
\label{eq:Lag}
\delta\mathcal{L}_{\rm EW} = c_{\phi}\mathcal{O}_{\phi} + c_{B} \mathcal{O}_B + c_W \mathcal{O}_{W} \, ,
\end{equation}
where $c_i$ stand for the Wilson coefficients of each operator:
\begin{align}
\mathcal{O}_{\phi} & = i \frac{\partial^\mu a }{f_a}\phi^\dagger \overleftrightarrow{D}_\mu \phi\, , \nonumber \\
\mathcal{O}_B & = - \frac{a}{f_a}B_{\mu\nu}\widetilde{B}^{\mu\nu} \, , \label{eq:operators} \\
\mathcal{O}_W & = - \frac{a}{f_a}W^I_{\mu\nu}\widetilde{W}^{\mu\nu}_I \, . \nonumber
\end{align}
Here, $\phi$ is the Higgs doublet while $B$ and $W^I$ stand for the EW vector bosons, and $a$ is the ALP field. The dual field strengths are defined as $\widetilde{X}^{\mu\nu} \equiv \frac{1}{2}\epsilon^{\mu\nu\rho\sigma}X_{\rho\sigma}$, with $\epsilon^{0123} = 1$, and $\phi^\dagger \overleftrightarrow{D}_\mu \phi \equiv \phi^\dagger \big{(}D_\mu\phi\big)-\big{(}D_\mu \phi\big{)}^\dagger \phi {}$.

It can be shown that the operator $\mathcal{O}_{\phi}$ can be eliminated through a hypercharge rotation~\cite{Georgi:1986df}
\begin{equation}
\begin{array}{rcl}
\phi & \to & e^{i c_\phi \frac{a}{f_a} }\phi \, ,\\
\Psi_F & \to & e^{2 i Y_F c_\phi \frac{a}{f_a}} \Psi_F \, ,
\end{array}
\end{equation} 
where $F$ runs over left- and right-handed quarks and charged leptons (with hypercharge $Y_F$). This trades the $\mathcal{O}_\phi$ operator by a set of fermionic operators of the form:
\begin{equation}
\label{eq:rotation}
 {\partial_\mu a(x)\over f_a}\sum_{F}~ \bar{\Psi}_F \gamma^\mu \Psi_F \, .
\end{equation}

\subsection{ALP production in kaon decays}

While the $\mathcal{O}_{\phi}$ and $\mathcal{O}_W$ operators are flavour-blind, they can lead to flavor-changing neutral-current (FCNC) processes via the exchange of gauge bosons at one loop, as discussed in detail in Refs.~\cite{Izaguirre:2016dfi,Gavela:2019wzg}. This will induce new kaon decay modes (\eg $K \to \pi a$).

We are interested in these hadronic processes at energies below the EW scale. The relevant couplings are the induced ALP couplings to the light quark currents in Eq.~\eqref{eq:rotation} 
\begin{equation}
{\partial_\mu a(x)\over f_a} \left(\sum_{q}~\bar{q}_{R} ~k_{q}  \gamma^\mu  q_{R} + \sum_{Q}  \bar{Q}_L ~k_Q\gamma^\mu Q_L\right) \, ,
\end{equation}
where following the notation of Ref.~\cite{Bauer:2021wjo}, lower and upper case for the quark field refers to the right-handed/left-handed quarks respectively, and $k_q$ and $k_Q$ are $3\times 3$ matrices with indices $(u,d,s)$. 
The flavour-diagonal axial-current couplings are\footnote{Here we neglect the differences between the couplings in the flavor and mass bases as they will only introduce subleading corrections in our case (we refer the interested reader to Ref.~\cite{Bauer:2020jbp} for a detailed discussion).}
\begin{eqnarray}
{\partial_\mu a(x)\over 2 f_a} \sum_{q=u,d,s}  c_{qq}~  \bar{q}~\gamma^\mu \gamma^5 q \, , 
\quad \mathrm{with} \quad 
c_{qq} \equiv [k_q -k_Q]_{qq} \, . 
\end{eqnarray}

The standard approach to incorporate these non-standard interactions in hadronic physics is to match the theory to Chiral Perturbation Theory ($\chi$PT) as first proposed in Ref.~\cite{Georgi:1986df}. Recently, this approach has been revisited and some previous inconsistencies related to weak hadron decays have been corrected \cite{Bauer:2021wjo}, in particular for the important decay $K \to \pi a$. The result for the amplitude found in Ref.~\cite{Bauer:2021wjo} is
\begin{align}
i\mathcal{A}_{K\to\pi a} = - & \frac{m_K^2 - m_\pi^2}{2f_a} [k_q+ k_Q]_{ds}  \nonumber \\
+ \frac{N_8}{4f_a} & \left\{ 6\left[c_{uu}+ c_{dd} - 2 c_{ss}\right] \frac{m_a^2 (m_K^2 - m_a^2)}{4m_K^2 - m_\pi^2 - 3m_a^2} 
\right. \nonumber \\
& + \left[2 c_{uu} +  c_{dd}+  c_{ss}\right] (m_K^2 - m_\pi^2 - m_a^2) + 4 c_{ss} m_a^2 \nonumber \\
& \left. + \left( [k_q+k_Q]_{dd} - [k_q + k_Q]_{ss} \right) (m_K^2 + m_\pi^2 - m_a^2)\right\} \, ,
\label{eq:Akpi}
\end{align}
where $m_a, m_\pi, m_K$ stand for the ALP, pion and kaon masses, respectively, and all the couplings in this amplitude are assumed at an energy scale $\mu=2$~GeV. The first term results from the induced flavor-changing vector currents in the EFT at low energies, while the rest of the terms are proportional to $N_8 \simeq - g_8 G_F/\sqrt{2} V_{us} V_{ud}^* f_\pi^2 \sim \mathcal{O}(1.5 \times 10^{-7})$, which is the dominant coupling of the weak Hamiltonian in $\chi$PT that mediates the standard ($\Delta I=1/2$) $K\rightarrow \pi\pi$ decays (see Refs.~\cite{Bauer:2021wjo, Cirigliano:2011ny} for details).

Next we need to connect the couplings $k_q(\mu), k_Q(\mu), c_{qq}(\mu)$ to the high-energy parameters, $c_\phi(\Lambda), c_W(\Lambda)$ and $c_B(\Lambda)$ in Eq.~(\ref{eq:Lag}). At tree level the only contribution comes from the operator ${\mathcal O}_\phi$, leading to flavour-diagonal and universal couplings of the form:
\begin{equation}
k_u(\Lambda)= -{4 \over 3} c_\phi(\Lambda), \quad  k_d(\Lambda)= {2 \over 3} c_\phi(\Lambda),  \quad k_Q(\Lambda) = -{1\over 3} c_\phi(\Lambda),
\end{equation}
for quarks and 
\begin{equation}
k_e(\Lambda) = 2 c_\phi(\Lambda), \quad k_L(\Lambda) =  c_\phi(\Lambda) ,  
\end{equation}
for charged leptons. This implies 
\begin{align}
\label{eq:cff}
c_{ff}(\Lambda)=\left\{
\begin{array}{ll} 
-c_\phi(\Lambda)\, \quad & \mathrm{for} ~ f=u,c,t \, , \\
c_\phi(\Lambda)\,  \quad & \mathrm{for} ~ f=d,s,b \, , \\
c_\phi(\Lambda)\,  \quad & \mathrm{for} ~ f=e,\mu,\tau \, .
\end{array}\right.
\end{align}
The axion operators mix under renormalization and, at the weak scale $\mu_{w}$, flavour non-diagonal contributions result from the matching to the low-energy EFT (\ie without $W,Z,t$). Let us stress that that the normalization used for the operators in Ref.~\cite{Bauer:2021wjo} differs from the one in Eq.~\eqref{eq:Lag}. Denoting the former as $\tilde{c}_i$, they are related as
\begin{eqnarray}
 {\alpha_2(\Lambda)\over 4\pi}  \tilde c_W(\Lambda) \equiv  -c_W(\Lambda), ~~{\alpha_1(\Lambda)\over 4\pi} \tilde c_B(\Lambda) \equiv  -c_B(\Lambda) \, , \label{eq:ctilde}
\end{eqnarray}
with $\alpha_1 \equiv \alpha / c_w^2$, $\alpha_2 \equiv \alpha / s_w^2$, where $s_w^2 \equiv \sin^2\theta_w \simeq 0.231$ is the Weinberg angle and $c_w^2 = 1 - s_w^2$. 
Setting $\mu_w$ to the top mass ($m_t$), the effective coupling $[k_Q]_{ds}$ reads~\cite{Bauer:2021wjo}
\begin{equation}
\label{eq:kdij}
[k_Q(m_t)]_{ds} = V_{td}^* V_{ts}  \left\{ -\frac{1}{6} I_t(m_t, \Lambda) + 
\frac{y_t^2(m_t)}{16\pi^2}c_{tt}(m_t)f(x_t) - \frac{y_t^2(m_t)}{16\pi^2} \frac{3 \alpha(m_t)}{2\pi s_w^2}\tilde c_{W}(\Lambda) h(x_t) \right\}
 \, ,
\end{equation}
where
\begin{equation}
h(x) \equiv \frac{1 - x + x \log x}{(1 - x)^2}, ~ 
f(x) \equiv -\frac{1}{2}\left[ \frac{1}{2} +  3 h(x)\right]\, ,
\end{equation}
while $V_{qq'}$ stands for the CKM matrix elements, $y_t$ is the top-quark Yukawa, $x_t \equiv m_t^2/m_W^2$ ($m_W$ being the mass of the $W$), and $\alpha$ is the fine structure constant. The function $I_t$ and the top axial current $c_{tt}$ depend on the couplings at $\Lambda$ as follows:
\begin{eqnarray}
I_t(\muew, \Lambda) &=&  - \tilde c_{W}(\Lambda)I_1(\muew,\Lambda) -\tilde c_{B}(\Lambda) I_3(\muew,\Lambda) \nonumber\\
&-& c_{\phi}(\Lambda) \left[ I_5(\muew,\Lambda) -\frac{2}{3} A(\muew, \Lambda)\right],
\label{eq:It}
\end{eqnarray}
and
\begin{eqnarray}
c_{tt}(\muew) & = & \tilde c_{W}(\Lambda)  I_2(\muew,\Lambda) + \tilde c_{B}(\Lambda) I_4(\muew,\Lambda)\nonumber\\
 &-&  c_{\phi}(\Lambda) \left[1 - A(\muew, \Lambda) - I_6 (\muew,\Lambda) \right] \, , \label{eq:ctt-run} 
\end{eqnarray}
where the integrals $I_{1-6}$ and the function $A$  are given in App.~\ref{app:integrals}. 

Substituting Eqs.~\eqref{eq:It} and~\eqref{eq:ctt-run} into Eq.~\eqref{eq:kdij} we obtain the following expression  in terms of the high-energy parameters:
\begin{align}
\label{eq:R}
\frac{[k_Q(\mu_w)]_{ds}}{V_{td}^*V_{ts}} =  \tilde c_{W}(\Lambda) & \left[ \frac{1}{6}I_1 + \frac{\alpha_t(\muew)}{4\pi} f(x_t) I_2  - \frac{\alpha_t(\muew)}{4\pi} \frac{3 \alpha(\muew)}{2\pi s_w^2} h(x_t)
\right] +\nonumber  \\
+ \tilde c_{B}(\Lambda) & \left[ \frac{1}{6}I_3+ \frac{\alpha_t(\muew)}{4\pi} f(x_t) I_4
\right]\nonumber  \\
- c_{\phi} (\Lambda)& \left[ -\frac{1}{6}I_5  + \frac{1}{9}A(\muew, \Lambda ) + 
\frac{\alpha_t(\muew)}{4\pi} f(x_t) \left[1 - A(\muew, \Lambda) - I_6  \right] \, , 
\right]
\end{align}
where $\alpha_t \equiv y_t^2/(4\pi)$. We are now in a position to evaluate the effective coupling numerically. Setting $\Lambda = 1~\rm{TeV}$, and substituting Eqs.~\eqref{eq:ctilde} into Eq.~\eqref{eq:R} we find
\begin{align}
\label{eq:keff-value}
\frac{[k_Q(\mu_w)]_{ds}}{V_{td}^*V_{ts}}\bigg|_{\Lambda=\rm{1 TeV}} \simeq   -9.7\times 10^{-3}c_W(\Lambda)  + 8.2\times 10^{-3}c_{\phi}(\Lambda) - 3.5\times 10^{-5}c_{B}(\Lambda) \, .
\end{align}
From this expression we see that the contribution from the $\mathcal{O}_B$ operator to the effective coupling is strongly suppressed, so the production mechanism will be dominated by the $\mathcal{O}_\phi$ and $\mathcal{O}_W$ operators. Furthermore Eq.~\eqref{eq:keff-value} is the leading contribution to the amplitude,  Eq.~\eqref{eq:Akpi}, being much larger than $N_8 \sim 10^{-7}$. No other flavour-changing coupling is generated, in particular $[k_q(\muew)]_{ds} = 0$. Moreover, once the EW bosons have been integrated out of the theory there is no additional running of the coupling constants, and we can therefore take $[k_Q (\muew)]_{ds}$ at energies below the EW scale, as discussed in Ref.~\cite{Bauer:2020jbp}. 
Hence to a good approximation we have
\begin{equation}
\label{eq:Akpi-approx}
i\mathcal{A}_{K\to\pi a} \simeq - \frac{m_K^2 - m_\pi^2}{2f}[ k_Q (\muew)]_{ds}  \, .
\end{equation}
The decay width obtained from this amplitude reads
\begin{align}
\label{eq:BKpia}
\Gamma(K^+\to \pi^+ a) &=\dfrac{m_{K}^3 \big|[k_Q(\muew)]_{sd}\big|^2}{64\pi f_a^2} f_0(m_a^2)\, \lambda_{\pi a}^{1/2}\left(1-\dfrac{m_{\pi}^2}{m_{K}^2}\right)^2 \,,
\end{align}
where $f_0$ denotes the scalar form factor\footnote{Within the mass range of interest here, it can be safely approximated as $f_0(q^2)\simeq 1$, see Ref.~\cite{Carrasco:2016kpy}.}, and we have defined 
\begin{align}
\lambda_{\pi a} & \equiv \lambda(1, m_a^2/m_K^2, m_\pi^2/m_K^2) \, , \\
\lambda(a,b,c) & = a^2 + b^2 + c^2 - 2ab - 2ac - 2bc \, .
\end{align}

Other recent estimates of this width (including tree-level and loop corrections) have been presented in Refs.~\cite{Izaguirre:2016dfi,Gavela:2019wzg, Guerrera:2021yss}. In Refs.~\cite{Izaguirre:2016dfi,Gavela:2019wzg}, the one-loop correction to $[k_Q]_{ds}$ from the operators ${\mathcal O}_W$ and ${\mathcal O}_\phi$ is assumed to be the dominant contribution. The physical amplitude mediated by this coupling is written in terms of the scalar form factor between a kaon and pion, and extracted from lattice QCD (see Refs.~\cite{Izaguirre:2016dfi,Gavela:2019wzg} for further details). This result is equivalent to that obtained in $\chi$PT as long as the effective coupling $[k_Q]_{ds}$ is the same, and if the form factor is set to one\footnote{In the case of $B\to K a$ decays, the decay width can be obtained from Eq.~\eqref{eq:BKpia} replacing the CKM matrix elements and meson masses by the corresponding ones in this case, and noting that the relevant form factor is $f_0^{B\to K}(q^2=0)\simeq 0.37$ (see \eg Fig.~33 in Ref.~\cite{Aoki:2021kgd}).}, which is a good approximation. The result quoted in Ref.~\cite{Gavela:2019wzg} for the effective coupling is
\begin{align}
\label{eq:R-approx}
\frac{[k_Q(\muew)]_{ds}}{V_{td}^*V_{ts}}\bigg|_{\Lambda=\rm{1 TeV}}  \simeq -{g_2(\muew)^2\over 16\pi^2} \left(3 c_{W}(\Lambda) g(x_t) -{c_\phi(\Lambda)
\over 4} x_t \log {\Lambda^2\over \muew^2}\right),
\end{align}
where $g(x) = x h(x)$, and $g_2(\muew)$ is the weak gauge coupling evaluated at the weak scale. 
Using Eq.~(\ref{eq:ctilde}) and the relations  $m_t = y_t  v / \sqrt{2}, \alpha_2 = \alpha/s_w^2, v = 2 m_W / g_2$, it is straightforward to show that Eq.~(\ref{eq:R-approx}) is  equivalent to the dominant contributions in Eq.~\eqref{eq:R}  up to a global sign:
\begin{align}
\label{eq:R-approx}
\frac{[k_Q]_{ds}}{V_{td}^*V_{ts}}\bigg|_{\Lambda=\rm{1 TeV}}  \simeq -\tilde c_{W}(\Lambda) & \left[  \frac{\alpha_t(\muew)}{4\pi} \frac{3 \alpha}{2\pi s_w^2} g(x_t)
\right] 
- c_{\phi}(\Lambda) \left[ \frac{1}{9}A(\muew, \Lambda )  \right],
\end{align}
since (see App.~\ref{app:integrals})
\begin{equation}
\label{eq:A-approx}
A(\mu, \Lambda) \simeq {9g_2^2\over 64\pi^2} x_t \log {\Lambda^2\over \mu^2} \, .
\end{equation}
Finally, in Ref.~\cite{Guerrera:2021yss} the authors considered additional contributions to $K\to\pi a$ at tree level, stemming from a non-universal ALP coupling to fermions. Even in that case the one-loop contributions are found to dominate over the tree-level diagrams, as long as the ALP coupling to the top quark in the loop is non-vanishing. It should also be noted that both the tree-level and one-loop diagrams (plus additional contributions not considered in from Ref.~\cite{Guerrera:2021yss}) are accounted for in the calculation using $\chi$PT, Eq.~\eqref{eq:Akpi}.

\subsection{ALP decay channels}
\label{sec:ALPdecay}

For ALP masses below 400~MeV, the decay channels that are kinematically open are $a\to \gamma\gamma$, $a\to e^+e^-$ and $a\to\mu^+ \mu^-$. The decay width into leptons can be written as:
\begin{equation}
\Gamma (a\to \ell^+\ell^-) =  
|c_{\ell\ell}|^2\dfrac{ m_a m_\ell^2 }{8\pi f_a^2} \sqrt{1-\dfrac{4 m_\ell^2}{m_a^2}} \, ,
\label{eq:GammaLepton}
\end{equation}
where at low energies ($\mu \sim 2~\mathrm{GeV}$) $c_{\ell\ell}$ is given at one loop by~\cite{Gavela:2019wzg,Bauer:2017ris}
\begin{align}
\begin{split}
c_{\ell\ell} = c_{\phi}&+\frac{3\,\alpha}{4\pi} \left(\frac{3\,c_W }{s_w^2}+ \frac{5\,c_B}{c_w^2 } \right) \log \dfrac{f_a}{m_W} 
+ \dfrac{6\, \alpha}{\pi}\left(c_B \, c_w^2+c_W\, s_w^2\right) \log \dfrac{m_W}{m_\ell} \,,
\end{split}
\label{eq:cll}
\end{align}
and to simplify the notation we have written $c_i \equiv c_i(\Lambda)$. 
Similarly, the decay width into two photons reads 
 \begin{align}
\label{eq:GammaPhoton}
\Gamma (a\to \gamma\gamma)& =  |c_{\gamma\gamma}|^2 \dfrac{m_a^3 }{4\pi f_a^2}\,,
\end{align}
where the effective coupling at low energies is given at one loop by~\cite{Gavela:2019wzg,Bauer:2017ris}
\begin{align}
\begin{split}
c_{\gamma \gamma } =  &c_W\,\Big[s_w^2\,+\frac{2\,\alpha}{\pi} B_2(\tau_W)\Big] +c_B\,c_w^2 
- c_{\phi} \,\frac{\alpha}{4\pi}\,\bigg( B_0 
- \frac{m_a^2}{m_{\pi}^2-m_a^2}\bigg)\, .
\end{split}
\label{eq:cgg}
\end{align}
Here, $B_0$ and $B_2$ are loop functions (which can be found in App.~\ref{app:loop}), and $\tau_W = 4m_W^2/m_a^2$.

A comparison of the different terms entering the effective couplings $c_{\gamma\gamma}$ and $c_{\ell\ell}$ allows to see that:
\begin{itemize}
\item the partial width $\Gamma(a\to\gamma\gamma)$ depends exclusively on the mass of the ALP, while $\Gamma(a\to \ell^+\ell^-)$ depends also on the lepton mass. This means that, for similar values of the effective couplings $c_{\gamma\gamma}$ and $c_{\ell\ell}$, then $\Gamma(a\to\gamma\gamma) \gg \Gamma(a\to \ell^+ \ell^-)$;
\item once the decay channel $a\to\mu^+\mu^-$ is open, it will completely dominate over the decay channel $a\to e^+ e^-$, due to the much larger muon mass; 
\item if $c_\phi \gg c_W,c_B$ the ALP will predominantly decay into lepton pairs since the terms in $c_{\ell\ell}$ which are proportional to $c_B, c_W$ are suppressed by $\alpha$. Conversely, if $c_W \gg c_\phi$ then the ALP will decay mostly into photons;
\item if loop-corrections are neglected, the effective coupling for $\Gamma(a\to\gamma\gamma)$ reads $c_{\gamma\gamma} \sim s_w^2 c_W + c_w^2 c_B $. Thus, the operator $\mathcal{O}_B$ can have a significant impact on (and even lead) the decay width for this channel, since $s_w^2 < c_w^2$.
\end{itemize}

Before concluding this section it is worth pointing out how the scenarios studied in this work may be mapped onto phenomenological \emph{benchmarks} commonly used in the literature, such as the ones defined in \eg Refs.~\cite{Beacham:2019nyx,Agrawal:2021dbo}. For the gluon-dominance benchmark there is no equivalence with our study, since we do not include the $G\tilde{G}$ operator. However, if the only operator included is $\mathcal{O}_W$, our results can be exactly mapped onto the so-called photon-dominance benchmark, since the ALP couples predominantly to photons in this case. The relation between the standard coupling used in this benchmark (see \eg Ref.~\cite{Beacham:2019nyx}) and $c_W$ is
\begin{equation}
\label{eq:gag}
g_{a\gamma} = \frac{c_W}{f_a} \left(4 s_w^2 + \frac{\alpha}{2\pi}\mathcal{O}\left(\frac{m_a}{m_W}\right)\right)\, .
\end{equation}
Finally, the case when only the $\mathcal{O}_\phi$ operator is included is \emph{approximately} equivalent to the fermion-dominance benchmark: while in our scenario the couplings to all fermions are not universally generated (see Eq.~\eqref{eq:cff}), our results in this case are fully dominated by the coupling to the top quark. Therefore the equivalence is realized up to a very good approximation. The corresponding relation is 
\begin{equation}
\label{eq:gY}
g_Y \simeq c_\phi \frac{v}{f_a} \, .
\end{equation}
When showing our results, we will also present the limits on the effective parameters $g_{a\gamma}$ and $g_Y$ for completeness.

\section{Previous constraints}
\label{sec:bounds}

In this section we provide a brief summary of the most relevant constraints for ALPs in the mass range $2 m_e < m_a < m_K - m_\pi$, and explain how these have been recasted to the specific scenario studied in this work. For a recent review of current bounds on ALPs in this mass region, see Ref.~\cite{Goudzovski:2022vbt} (and references therein, such as Refs.~\cite{Dolan:2017osp, Dobrich:2019dxc, Essig:2010gu, Bauer:2017ris, Bauer:2021mvw, Izaguirre:2016dfi, Jaeckel:2015jla}). 

\subsection*{Visible ALP decays} At fixed-target experiments, the ALP can be produced through Primakoff scattering (that is, the conversion of a photon into an ALP in the vicinity of a nucleus), or through its mixing with pseudoscalar mesons produced in the target ($\pi^0,\eta,\eta'$). If it couples directly to quarks or electrons, it can also be produced by proton or electron Bremsstrahlung. The analyses performed in the literature typically assume that the ALP is coupled predominantly to either photons or electrons and therefore the leading production and decay mechanisms will be different in the two scenarios. If the ALP predominantly couples to photons, the strongest constraints below 500 MeV come from the E137 experiment~\cite{Bjorken:1988as}, see \eg Refs.~\cite{Dolan:2017osp,Dobrich:2019dxc,Essig:2010gu}. We take the limit from Ref.~\cite{Dolan:2017osp}, where the E137 bound was recasted on the ALP coupling to photons after EW symmetry breaking, $g_{a\gamma\gamma}$. Since they were obtained assuming that the ALP couples predominantly to photons, they can be directly applied\footnote{The bound is rescaled as $g_{a\gamma\gamma}/4 = c_W s_w^2$, considering the different normalization used in Ref.~\cite{Dolan:2017osp} for the $\mathcal{O}_W$ operator.} to the $c_W$ coupling. Conversely, if $\mathcal{O}_\phi$ dominates the ALP couples mostly to electrons. In this case we use the analysis of Ref.~\cite{Essig:2010gu}, which computed the  excluded regions for E137 and CHARM~\cite{CHARM:1985anb} assuming the ALP is predominantly produced through its mixing with pseudoscalar mesons, and that it decays into $e^+e^-$ pairs. Here we use their CHARM bound, which is stronger than the one obtained from E137. 

At higher masses, significant constraints are also obtained from CHARM~\cite{CHARM:1985anb} and LHCb~\cite{LHCb:2015nkv,LHCb:2016awg}, assuming that the ALP is produced from $B$-meson decays (via $B \to K a$) and that it decays visibly within the detector as $a\to \mu^+ \mu^-$. Note that, although the obtained constraints in this case are less stringent they are applicable in a wider mass range, $2 m_\mu < m_a < m_B - m_K$. We take these from Refs.~\cite{Dobrich:2018jyi,Gavela:2019wzg}. Relevant bounds are also obtained from searches for $B\to K a,~a\to\gamma\gamma$ at BaBar~\cite{BaBar:2021ich}.

Additional bounds can be obtained from NA64, where the ALPs would produced in the forward direction through the Primakoff effect in interactions of high-energy Bremsstrahlung photons with nuclei in the target: $e^- N(A,Z) \to e^- N(A,Z) \gamma$, followed by $\gamma N(A,Z) \to a N(A,Z)$ where the ALP is produced through the exchange of a virtual photon with the nucleus $N(A,Z)$. The collaboration reported a limit for the search of ALPs decaying into two photons in Ref.~\cite{NA64:2020qwq}. Finally, at electron-positron colliders ALPs could be produced through an off-shell photon (for example, via $e^+ e^- \to \gamma^* \to \gamma a$) or in photon fusion ($e^+e^- \to e^+ e^- a$) and decay to two photons. The strongest bound in this case is obtained from a reanalysis of LEP data~\cite{Jaeckel:2015jla}. In our scenario, both NA64 and LEP bounds apply to the $c_W$ coupling. 

\subsection*{Invisible ALP decays.} If the ALP is sufficiently long-lived (for very light masses, or small enough couplings) it may exit the detector without decaying. This would lead to an excess of decays of kaons into pions plus missing energy ($K \to \pi + \mathrm{inv}$), or $B$ mesons to kaons plus missing energy ($B \to K + \mathrm{inv}$), which can be constrained using data from $K \to \pi \nu\bar\nu$ or $B \to K \bar\nu \nu$ searches~\cite{Essig:2010gu,Izaguirre:2016dfi,Gavela:2019wzg}. In this work, unlike in Refs.~\cite{Izaguirre:2016dfi,Gavela:2019wzg} we do not assume significant ALP couplings to the dark sector and, therefore, we need to take into account that if the ALP is sufficiently short-lived it will not contribute to the mentioned observables. Thus, an upper bound is imposed on $\text{BR}(M\to M^\prime a) \times P_\text{exit}$, where $M,M'$ stand for the parent and daughter mesons, respectively, and $P_\text{exit} \sim e^{-\Delta \ell_\mathrm{det} /L_a}$ is the probability of the ALP to exit the detector without decaying. Here $\Delta \ell_\mathrm{det}$ stands for the approximate detector size and $L_a \equiv \gamma_a \beta_a c\tau_a$ is the ALP decay length in the lab frame ($\tau_a$ denotes the proper lifetime of the ALP and $\gamma_a, \beta_a$ correspond to the boost variables, while $c$ is the speed of light).

The strongest limits on $K\to\pi a$ are obtained from the NA62 experiment~\cite{NA62:2020xlg,NA62:2020pwi}, where we assume $\Delta \ell_\text{det} \simeq 100\; \text{m}$ as detector size and $p_K \simeq 75~\mathrm{GeV}$ as the kaon momentum~\cite{NA62:2020pwi}. However, the very competitive bounds from E787 \& E949~\cite{BNL-E949:2009dza} are comparable (and even dominate) in the region close to the pion mass. In this case we assume the kaon decays at rest, and we take $\Delta \ell_\text{det} \sim 1.5~\mathrm{m}$. In the case of $B \to K a$ decays, we saturate the upper bound from the Belle experiment on $B^+ \to K^+ \nu\bar\nu$~\cite{Belle:2017oht}, taking $\Delta \ell_\mathrm{det} \simeq 5\ \mathrm{m}$~\cite{Belle:2012iwr} and $p_a \simeq 2.5\ \mathrm{GeV}$. Experimental limits on invisible ALP decays are also obtained from precision measurements of the pion momentum in $K \to \pi X$: if a two-body decay of the form $K\to\pi a$ takes place, this would lead to a monochromatic line in the pion momentum distribution. We take the limit from Ref.~\cite{Yamazaki:1984vg}: since no photon veto is imposed in this analysis, this bound always applies regardless of the lifetime of the ALP.

In principle, searches for mono-photon signals at colliders (\eg at BaBar~\cite{BaBar:2008aby,BaBar:2017tiz} or LEP~\cite{L3:1997exg}) may be reinterpreted in the context of an ALP that decays outside the detector~\cite{Izaguirre:2013uxa,Essig:2013vha,Izaguirre:2016dfi}. However, these are milder than the rest of the limits considered in this work.

\subsection*{ALP contributions to kaon three-body decays.} 

A third set of constraints can be derived from the decays $K\to\pi ee$ or $K\to\pi \gamma\gamma$. An ALP would contribute to these through the same penguin diagram leading to $K \to \pi$, with an extra vertex coupling the ALP propagator (which can be either on-shell \cite{Izaguirre:2016dfi} or off-shell~\cite{Gavela:2019wzg}) to either two photons or an electron-positron pair. If the dominant coupling is $c_\phi$ these are superseeded by other constraints; however if $c_W$ dominates, relevant bounds are obtained from measurements of $K^\pm \to \pi^\pm \gamma\gamma$ at NA62~\cite{NA62:2014ybm} and E949~\cite{E949:2005qiy}, and $K_L \to \pi^0 \gamma\gamma$ at NA48/2~\cite{NA48:2002xke} and KTeV~\cite{KTeV:2008nqz}, which we take from Ref.~\cite{Goudzovski:2022vbt}.  

\subsection*{Astrophysical bounds.} 

Three main bounds are obtained from supernovae data: (1) from the requirement that the energy loss induced by ALP emission does not exceed the energy loss from neutrino emission~\cite{Raffelt:1990yz,Raffelt:1987yt}; (2) from the visible signal resulting from the ALP burst, in the case where the ALP decays into pairs of photons~\cite{Jaeckel:2017tud}; and (3) from the observation of low-luminosity core-collapse supernovae, which constrains the total energy deposition in the progenitor star from radiative ALP decays, such as $a\to\gamma\gamma$ or $a\to e^+ e^-$~\cite{Caputo:2022mah}. In this work we focus on laboratory experiments, and refer the interested reader to Refs.~\cite{Chang:2018rso,Croon:2020lrf,Caputo:2022mah} for recent updates on astrophysical constraints. We note that the supernova bounds explore a different region of parameter space than laboratory experiments, since they typically apply to smaller couplings and masses (see \eg Fig.~7 in Ref.~\cite{Goudzovski:2022vbt}).

\section{Axion-like particles at MicroBooNE}
\label{sec:results}

The results from Sec.~\ref{sec:chipt} can be used to compute the branching ratio into the $K \to \pi a$ channel as well as the differential angular distribution of the ALP flux. 

In order to compute the expected number of events at MicroBooNE from the NuMI absorber, we use the same NuMI kaon distributions as in Ref.~\cite{Coloma:2015pih}, which were derived from a Monte Carlo simulation of the NuMI target when exposed to a 120 GeV proton beam. As outlined in the introduction, the expected ALP flux from the NuMI absorber would correspond to that of kaons decaying at rest. Since the production takes place via a two-body decay an isotropic flux is expected, with energy $E_a = (-m_\pi^2 + m_a^2 + m_K^2)/(2 m_K)$. 

For a total number of $N_K$ kaon decays, the event rate expected from ALP decays into $e^+e^-$ pairs inside the MicroBooNE detector can be computed as
\begin{equation}
N_{events} = \frac{N_K \times \text{BR}(K \to \pi a)}{4\pi} \; \text{BR} (a \to e^+e^-) \; \epsilon_{eff}\int_{\Delta\Omega_\text{det}} d\Omega~ P_\text{decay} (\Omega) \, ,
\end{equation}
where the integral runs over all trajectories with solid angle $\Omega$ intersecting the detector, and $\Delta\Omega_\text{det}$ is the solid angle of the detector as seen from the absorber. Here $\epsilon_{eff}$ stands for the detection efficiency (which depends on $m_a$) and $\text{BR}(a\to e^+ e^-)$ is the branching ratio for the ALP to decay into an electron-positron pair. Moreover, we have assumed an isotropic flux of kaons as well as a point-like absorber. Finally, $P_\text{decay}$ represents the probability of an ALP to decay inside the detector: 
\begin{equation}
\label{eq:Pdecay}
P_\text{decay} = e^{-\frac{\ell_\text{det} }{L_a }} \left[ 1 - e^{-\frac{\Delta \ell_\text{det}}{L_a}} \right] \, ,
\end{equation}
where $\ell_\text{det}$ is the distance traveled before it reaches the detector, and $\Delta \ell_\text{det}$ is the length of the ALP trajectory intersecting the detector. In practice, both $\ell_\text{det}$ and $\Delta \ell_\text{det}$ depend on the angle of the ALP trajectory. In this work we perform a numerical integration over all ALP trajectories, using the same detector dimensions and orientation with respect to the absorber as in Ref.~\cite{MicroBooNE:2021usw}. However, for estimation purposes it is useful to neglect the dependence of $P_\text{decay}$ with the solid angle and write the expected number of events as
\begin{equation}
\label{eq:events}
N_{events} \sim N_K \times \text{BR}(K\to \pi a ) \times \varepsilon_\text{det} \times \epsilon_{eff} \times \langle P_\text{decay} \rangle \times \text{BR}(a \to e^+ e^-) \, ,
\end{equation}
where $\langle P_\text{decay} \rangle$ represents the average probability of an ALP to decay inside the detector (obtained assuming as typical values $\ell_\text{det} \sim 100$ m and $\Delta \ell_\text{det} \sim 4$ m), while $\varepsilon_\text{det} = \Delta \Omega_\text{det}/(4\pi) \sim 1.5 \times 10^{-3}$ stands for the detector geometric acceptance. Assuming that the $\mathcal{O}_\phi$ coupling dominates, the branching ratio into electrons dominates, $BR(a\to e^+e^-) \simeq 1$, see Sec.~\ref{sec:ALPdecay}. Regarding the total number of kaons produced, numerically we find $N_K \sim 0.2 N_\text{PoT}$ kaons produced in the NuMI absorber, where $N_\text{PoT}$ stands for the number of Protons on Target (PoT) considered. According to this estimate, we find that a few signal events are a priori expected for $m_a = 100$ MeV and $c_\phi \sim 2\times 10^{-3}$, for an exposure of $N_\text{PoT} \sim 2 \times 10^{20}$. As we will see below, this agrees very well with the results from the exact numerical calculation. 

The rest of this section is structured as follows. First, we present the current constraints derived from MicroBooNE data in Sec.~\ref{sec:present}. In Sec.~\ref{sec:future} we proceed and determine the future sensitivity to this scenario using the whole collected (unprocessed) data, and consider additional search channels in addition to $a\to e^+e^-$. Finally, in Sec.~\ref{sec:interference} we study possible interference effects when two operators are simultaneously included in the analysis.

\subsection{Bounds from current MicroBooNE data}
\label{sec:present}

In order to determine the limit on the effective Wilson coefficients, we follow the same approach as in Ref. \cite{MicroBooNE:2021usw} and perform an unbinned $\chi^2$ analysis, taking $n_{bg} = 1.9$ as the expected total number of background events and $n_\text{obs} = 1$ as the observed number of events\footnote{The collaboration observed two candidate events in the dataset passing the cuts; however, one of them was rejected.}. We also include two nuisance parameters in order to account for systematic uncertainties affecting the overall normalization of the events for the signal ($\xi_s$) and background ($\xi_b$). For each uncertainty a penalty term is added to the $\chi^2$, which is then minimized over the nuisance parameters. Our $\chi^2$ reads:
\begin{equation}
\label{eq:chi2}
\chi^2 = \text{min}_{\xi_s, \xi_b} \left\{ 2(n_\text{pred} - n_\text{obs}) + 2n_\text{obs} \log\frac{n_\text{obs}}{n_\text{pred}} \right\} + \frac{\xi_{s}^2}{\sigma_s^2} + \frac{\xi_{b}^2}{\sigma_b^2}  \, ,
\end{equation}
where we take $\sigma_s = 30\%$ for the signal and $\sigma_b = 89\%$ for the backgrounds (according to Tab.~I in Ref.~\cite{MicroBooNE:2021usw}). Here $n_\text{pred}$ is the predicted number of events 
\begin{equation}
\label{eq:pred}
  n_\text{pred}(\{c\}, m_a,\xi_s, \xi_b) = \left[1 + \xi_s\right] \, n_{sig}(\{c\}, m_a) + \left[1 + \xi_b\right] \, n_{bg} \, ,
\end{equation}
where $n_{sig}$ stands for the signal predicted by the model for an ALP with mass $m_a$ and a given set of operator coefficients $\{c\}$. When computing $n_{sig}$ we take the selection efficiencies provided in the supplemental material of Ref.~\cite{MicroBooNE:2021usw} as a function of the scalar mass. We believe these efficiencies are directly applicable to our model given that the final topology for the decay is exactly the same. 

\begin{figure}[ht!]
  \includegraphics[width=1.0\textwidth]{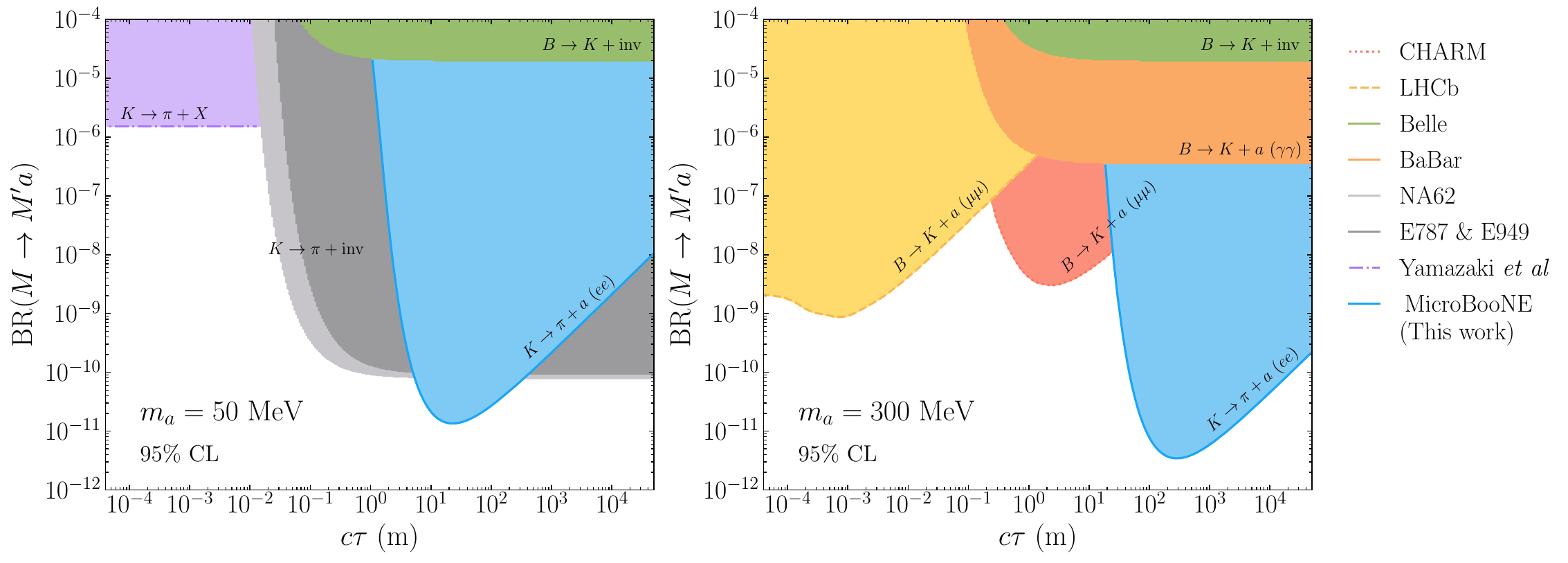}
\caption{\label{fig:br-ctau} MicroBooNE bounds (at 95\% C.L.) in the plane of $c\tau$ vs the ALP production branching ratio, for two different ALP masses as indicated by the labels (blue regions). Our bounds are obtained from a reanalysis of the data in Ref.~\cite{MicroBooNE:2021usw}, which used $1.93\times 10^{20}$ Protons on Target (PoT) and looked for $e^+ e^-$ pairs. For comparison we also show  the constraints (at 90\% CL) from invisible decays derived from Belle~\cite{Belle:2017oht}, NA62~\cite{NA62:2020xlg,NA62:2020pwi} and E787 \& E949~\cite{BNL-E949:2009dza} data, as well as the constraints (at 95\% CL) from searches for visible ALP decays from LHCb~\cite{LHCb:2015nkv,LHCb:2016awg} and CHARM~\cite{CHARM:1985anb,CHARM:1983ayi} (taken from Ref.~\cite{Dobrich:2018jyi}), and from BaBar data~\cite{BaBar:2021ich}. Note that the LHCb and CHARM constraints only apply to ALP masses $m_a > 2 m_\mu$ while the limits from NA62 and E787 \& E949 only apply for $m_a < m_K - m_\pi$. We also show the constraints from Yamazaki \emph{et al.}~\cite{Yamazaki:1984vg} which apply regardless of the ALP lifetime, see text for details. In this figure we only show constraints obtained from on-shell production of the ALP in meson decays. }
\end{figure}

Our derived constraints using the $\chi^2$ definition in Eq.~\eqref{eq:chi2} are presented in Fig.~\ref{fig:br-ctau}, under the assumption that the production branching ratio and the decay width are completely independent.
Although in a given model the two quantities are typically correlated, this allows us to derive a model-independent constraint so that our results can be easily recasted for other scenarios. Note that, for each line shown, the bound is presented assuming that the branching ratio of the decay into the channel indicated is 1; for a given model these would have to be rescaled according to the corresponding branching ratios. The MicroBooNE region (shown in blue) shows the portion of parameter space where we obtain a $\chi^2 > 3.84$, corresponding to 95\% confidence level (CL) for a counting experiment. Our results are shown for two illustrative values of the mass of the ALP, below (left panel) and above the kaon mass (right panel). 

\begin{figure}[ht!]
\begin{center}
  \includegraphics[width=0.95\textwidth]{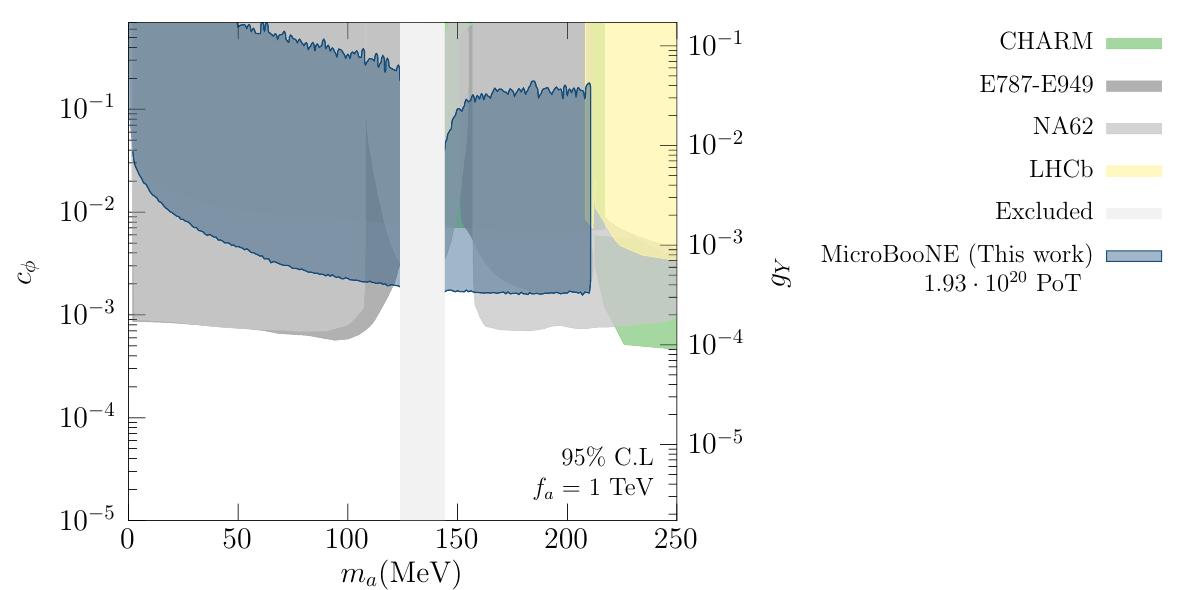}
\end{center}
\caption{\label{fig:micro-bound} MicroBooNE bounds (at 95\% C.L.) on $c_{\phi}$, for $f_a = 1 \; \text{TeV}$ and setting $c_W = c_B=0$. These are obtained from a reanalysis of the data in Ref.~\cite{MicroBooNE:2021usw}, which used $1.93\times 10^{20}$ Protons on Target (PoT). For comparison we also show the constraints from the NA62 experiment~\cite{NA62:2021zjw,NA62:2020xlg}, E787 \& E949~\cite{BNL-E949:2009dza}, CHARM~\cite{Essig:2010gu, CHARM:1985anb} and LHCb~\cite{LHCb:2015nkv,LHCb:2016awg, Gavela:2019wzg}. The shaded vertical band is excluded due to the large $a-\pi^0$ mixing, and is taken from Ref.~\cite{Gavela:2019wzg}. In the right axis we show the corresponding limit on the effective coupling for the so-called fermion-dominance scenario (defined \eg in Refs.~\cite{Beacham:2019nyx,Agrawal:2021dbo}), see Eq.~\eqref{eq:gY}. }
\end{figure}
In a given model, however, the values of the branching ratios and the lifetime of the ALP will be correlated since they both will depend on the same set of couplings as well as on the mass of the ALP. This typically leads to decrease in sensitivity for experiments searching for visible ALP decays, as is the case of MicroBooNE in this work. For example, from the left panel in Fig.~\ref{fig:br-ctau} we see that MicroBooNE would be in principle sensitive to values of the production branching ratio of the ALP at the level of $\mathcal{O}(2\times 10^{-11})$. Taking into account the expression for the branching ratio in Eq.~\eqref{eq:BKpia}, for a mass of 50 MeV this can be obtained setting $c_{W} \sim \mathcal{O}(5\times 10^{-4})$; however, this leads to very large values of $c\tau \sim 1500$~m (and even larger values if we assume that the ALP is coupled via the $\mathcal{O}_{\phi}$ operator instead of $\mathcal{O}_W$). In this regime the expected number of decays in the detector scales as $\Delta \ell_\text{det}/L_a$, see Eq.~\eqref{eq:Pdecay}, suppressing the sensitivity for very large values of $L_a$. 
Therefore, once we impose this restriction the relative comparison between the sensitivity of visible and invisible decay searches changes dramatically, as shown in Fig.~\ref{fig:micro-bound}. However, in spite of this we still find that MicroBooNE is generally competitive with current constraints, and is able to improve over these for ALP masses close to the pion mass.

\subsection{Future sensitivity}
\label{sec:future}

The limits shown in Sec.~\ref{sec:present} were obtained using only a reduced subset of the available NuMI data, and only searching for decays into $e^+ e^-$ pairs. However, according to Ref.~\cite{MicroBooNE:2021usw}, MicroBooNE has collected about 10 times more NuMI data, which has not been processed yet. Moreover, the MicroBooNE detector consists on a Liquid Argon Time Projection Chamber (LAr TPC), which allows to study not only final state topologies involving $e^+ e^-$ pairs, but also $\gamma\gamma$ and $\mu^+ \mu^-$ pairs. Here we compute the expected sensitivity using the full NuMI dataset recorded at MicroBooNE (corresponding to $\approx 2.2 \times 10^{21}$ PoT), and for three different searches using the three possible final states for the decay of the ALP. 

\begin{figure}[ht!]
\begin{center}
  \includegraphics[width=0.95\textwidth]{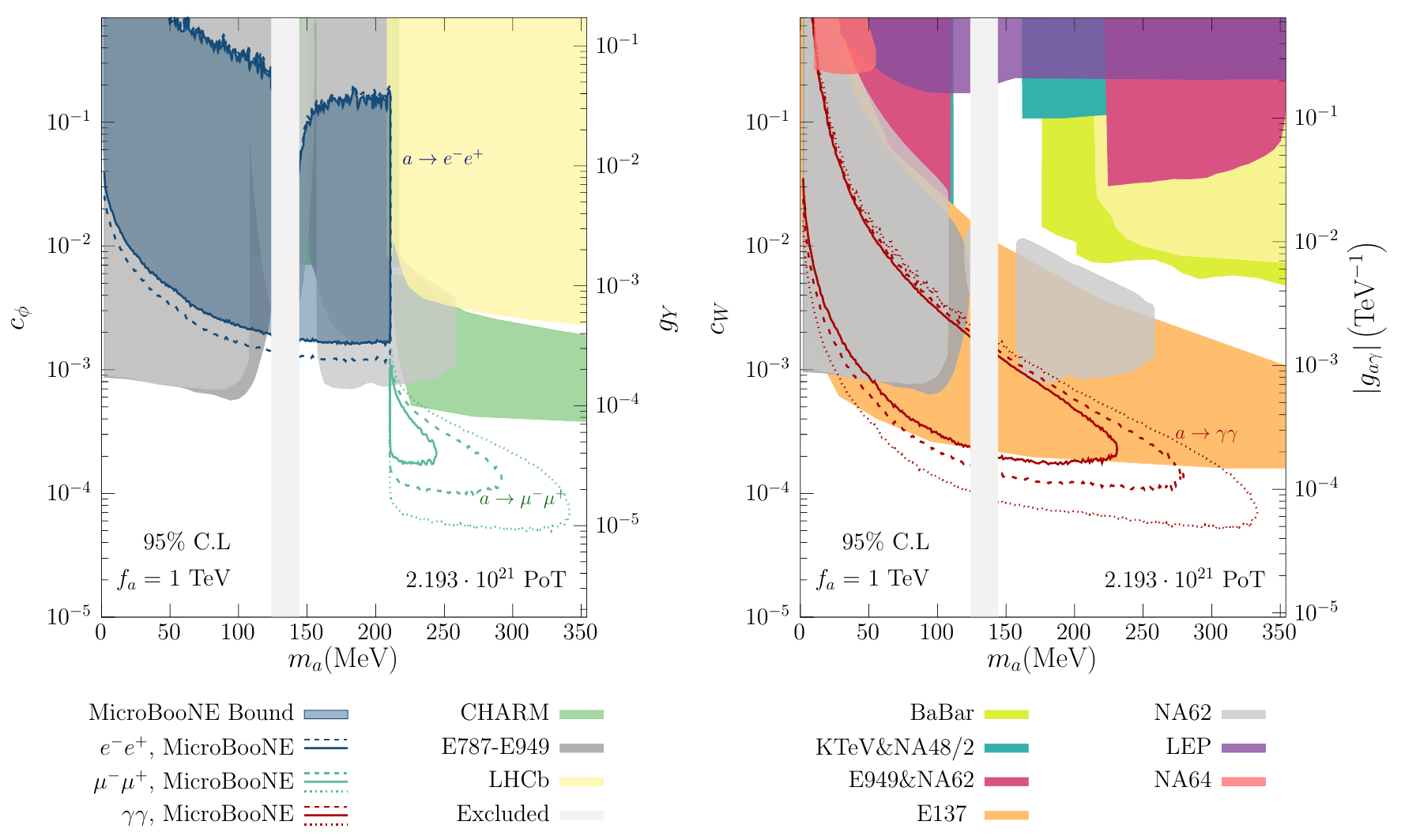}
\end{center}
\caption{\label{fig:micro-sens} MicroBooNE sensitivity projections, for $c_{\phi}$ (left panel) and $c_{W}$ (right panel) as a function of $m_a$, assuming $f_a=1\;\text{TeV}$, for $N_\text{PoT}=2.2\times 10^{21}$. The regions enclosed by the colored lines satisfy $\Delta \chi^2 > 3.84$, corresponding to 95\% CL for 1 degree of freedom (d.o.f.). Results are shown separately for the decay channels with the largest branching ratios in each case: $a \to \gamma\gamma$, $a\to e^+ e^-$ and $a\to \mu^+ \mu^-$, as indicated by the labels. Solid lines assume $\sigma_s=0.89$ and $\sigma_b=0.3$; dashed lines assume $\sigma_s=\sigma_b=0.2$; and dotted lines indicate the no-background limit, taking $\sigma_s=0.2$ (see text for details). The shaded areas show current bounds from BaBar~\cite{BaBar:2021ich}, E137~\cite{Bjorken:1988as,Dolan:2017osp}, NA62~\cite{NA62:2021zjw,NA62:2020xlg}, E787 \& E949~\cite{BNL-E949:2009dza}, LHCb~\cite{LHCb:2015nkv,LHCb:2016awg,Gavela:2019wzg}, CHARM~\cite{CHARM:1985anb,Essig:2010gu}, NA64~\cite{NA64:2020qwq} and LEP~\cite{Jaeckel:2015jla}. We also show bounds from measurements of $K\to\pi\gamma\gamma$ at E949~\cite{E949:2005qiy}, NA48/2~\cite{NA48:2002xke}, NA62~\cite{NA62:2014ybm} and KTeV\cite{KTeV:2008nqz} (taken from Ref.~\cite{Goudzovski:2022vbt}), as well as our bound derived from MicroBooNE data in Sec.~\ref{sec:present}. The shaded vertical band is disfavored due to the large $a-\pi^0$ mixing, and is taken from Ref.~\cite{Gavela:2019wzg}. In each panel only one operator is switched on at a time, setting the remaining operator coefficients to zero. The right axes show the corresponding limits on the effective couplings for the so-called photon-dominance and fermion-dominance scenarios, see Eqs.~\eqref{eq:gag} and~\eqref{eq:gY}. }
\end{figure}

The same $\chi^2$ definition as in the previous section, Eq.~\eqref{eq:chi2}, will be used here. However, in this case we assume that the experiment will observe a number of events fully consistent with the background expectation. In order to derive the limits on the couplings, we perform a hypothesis test, that is, for a fixed value of the mass we compute a $\Delta\chi^2$ as 
\begin{equation}
\Delta \chi^2(m_a, \left\{ c \right\}) = \chi^2(m_a, \left\{ c \right\}) - \chi^2_{\mathrm{SM}} \, .
\end{equation}
which is expected to be chi-squared distributed. The sensitivity regions are determined by the set of couplings for which we obtain $\Delta \chi^2(\left\{ c \right\})$ above a certain value. 

For the $ee$ channel, the number of background events is estimated rescaling the predicted background in Ref.~\cite{MicroBooNE:2021usw} by a factor of $\approx 10$ in order to account for the much larger exposure. Regarding the $\mu\mu$ and $\gamma\gamma$ channels the backgrounds may of course be very different, the same angular and timing cuts as for the $ee$ channel may also be applied here. In the case of the $\gamma\gamma$ channel, while in principle a neutrino NC$1\pi^0$ interaction could induce a background, the invariant mass of the two photons would reconstruct to the $\pi^0$ mass. In the case of the $\mu^-\mu^+$ channel, a possible background could arise from CC$1\pi$ neutrino events, where the charged pion is mis-reconstructed as a muon. However, additional cuts could potentially reduce this background further: for example, an analysis of a similar background at the DUNE near detector in the context of trident searches (where two muons are also expected for the signal) found that it could be reduced by seven orders of magnitude while keeping a very good signal efficiency~\cite{Altmannshofer:2019zhy}. Thus, we conservatively assume that the total background counts for the $\gamma\gamma$ and $\mu\mu$ channels will be similar in magnitude than for the $ee$ channel; however, we will also show the results obtained under the more optimistic assumption that the background can be reduced down to a negligible level. In all cases, we use the same same detection efficiencies as in Ref.~\cite{MicroBooNE:2021usw} up to $m_a = 210$ MeV, and a flat efficiency at 9.6\% for higher masses.

Figure \ref{fig:micro-sens} shows the expected sensitivity from a search using all recorded NuMI data at MicroBooNE. Our projections are shown for the ALP decay channels yielding the largest branching ratios in each case: $a \to e^+ e^-$ and $a \to\mu^+ \mu^-$ in the case of the $\mathcal{O}_\phi$ operator (left panel), and $a \to \gamma\gamma$ in the case of the $\mathcal{O}_W$ operator (right panel), see Sec.~\ref{sec:ALPdecay}. For comparison, the regions of parameter space disfavored by current experiments (including the MicroBooNE bound from Fig.~\ref{fig:micro-bound}) are indicated by the shaded areas, while the colored lines indicate the future sensitivity contours for MicroBooNE at 95\% CL (for 1 d.o.f.). Note that, while the $\mathcal{O}_\phi$ operator would lead to $\ell^+ \ell^-$ signals,  sensitivity to the $\mathcal{O}_W$ is possible only from a search in the  $\gamma\gamma$ channel since the contribution of this operator to $c_{\ell\ell}$ is suppressed by $\alpha$, see Eq.~\eqref{eq:cll}. This illustrates the importance of conducting searches with different final state topologies, making full use of the LAr TPC capabilities. Also, note the $\sim 10$ times better sensitivity to $c_W$ compared to $c_\phi$. Although the two couplings enter on equal terms at the production level, the decay width for $c_W$ is larger by roughly ${\mathcal O}(m_a^2/m_l^2)$ compared to that of $c_\phi$ leading to a shorter ALP lifetime. This eventually enhances the event rate in the detector, since in the limit of small couplings the decay probability can be approximated as $P_\text{decay} \propto \Delta \ell_\text{det} / L_a $, see Eq.~\eqref{eq:Pdecay}.

Due to the very large exposure considered, we find that the bound is no longer statistics-limited but limited by systematics instead. While the results in Ref.~\cite{MicroBooNE:2021usw} were obtained for very conservative systematic uncertainties, it may be possible to reduce these considerably. For example, their background uncertainties were dominated by the detector modeling (and, in particular, by the low statistics of simulated neutrino events passing selection cuts), while cross section uncertainties may be further reduced with dedicated studies and improvement of theoretical models.  Therefore, Fig.~\ref{fig:micro-sens} shows solid lines, corresponding to the same set of uncertainties as in Fig. \ref{fig:micro-bound} ($\sigma_s = 30\%,~ \sigma_b=89\%$); and dashed lines, corresponding to $\sigma_s = \sigma_b = 20\%$. While in the former case the improvement over the present MicroBooNE constraint for $c_\phi$ would only be marginal (left panel), from the comparison of the solid and dashed lines it can be seen that there is room for improvement provided that systematic uncertainties can be significantly reduced below current values. Additionally, our results show the potential to constrain an unexplored region of parameter space for $c_\phi$ by looking for ALP decays into $\mu^+\mu^-$ final states. For $c_W$ (right panel) we also find that the bounds from MicroBooNE would be competitive with those obtained from beam-dump experiments and may even be able to improve over these, depending on the level of systematic uncertainties assumed. Finally, given the dependence of the results on the background level assumed, the dotted lines indicate the expected sensitivity in the no-background limit, assuming that additional cuts could bring it down to a negligible level for the $\mu^+\mu^-$ and $\gamma\gamma$ channels (in both cases, 20\% signal systematics have been assumed).

\subsection{Interference effects between different operators}
\label{sec:interference}

\begin{figure}[ht!]
\begin{center}
  \includegraphics[width=0.85\textwidth]{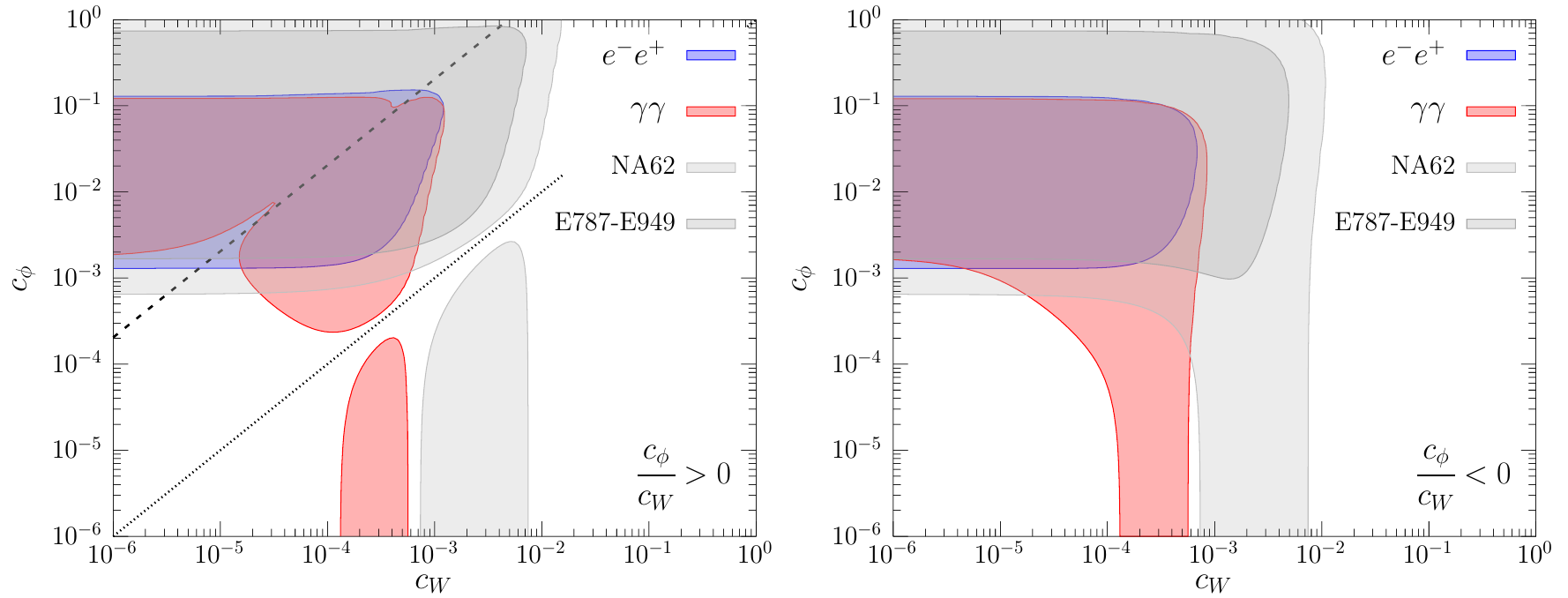}
  \includegraphics[width=0.85\textwidth]{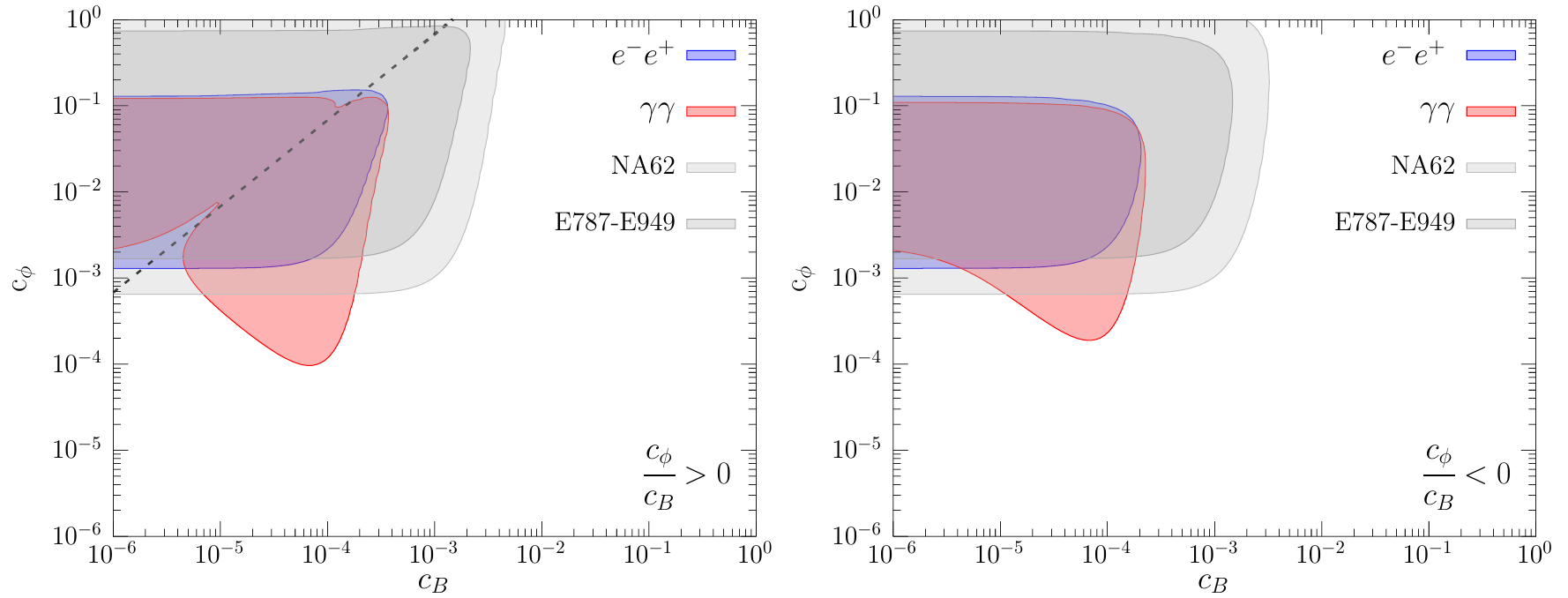}
  \includegraphics[width=0.85\textwidth]{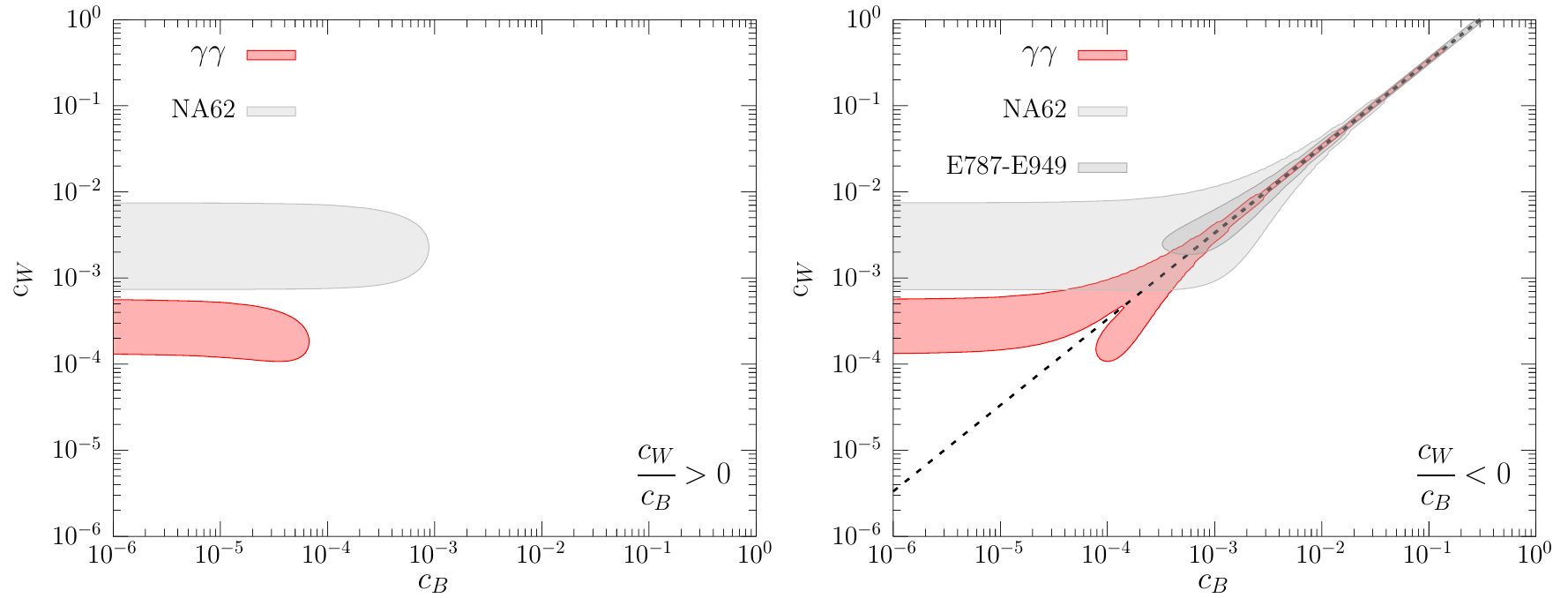}
\end{center}
\caption{\label{fig:cB} 
MicroBooNE sensitivity projections, assuming two couplings are non-zero at a time. The shaded colored regions satisfy $\Delta \chi^2 > 5.99$ (corresponding to the sensitivity at 95\% CL for 2 d.o.f.) from a search for a $\gamma\gamma$ (red) or $e^+e^-$ (blue) signal. Results are shown for $f_a=1\;\text{TeV}$ and $m_a = 200~\mathrm{MeV}$, for $N_\text{PoT}=2.2\times 10^{21}$, and assuming $\sigma_{s}=\sigma_{b}=20\%$. Left (right) panels are obtained assuming a positive (negative) relative sign between the operators introduced. 
The diagonal lines indicate destructive interference between different operators entering the production rate (dotted lines) or decay width (dashed lines), see text for details. In all panels, the shaded gray regions indicate the portions of parameter space disfavored by NA62~\cite{NA62:2021zjw,NA62:2020xlg} (light gray), and E787 \& E949~\cite{BNL-E949:2009dza} (dark gray). 
}
\end{figure}

So far our results have been obtained under the assumption that only one of the operators which control the production of the ALP ($\mathcal{O}_W$, or $\mathcal{O}_\phi$) is present in the effective Lagrangian and, for simplicity, the coefficient that goes with $\mathcal{O}_B$ has also been set to zero. However, it has been pointed out previously in the literature~\cite{Gavela:2019wzg} that destructive interference between different EW operators could significantly affect the effective couplings controlling both the ALP production and decay. Thus, in Fig.~\ref{fig:cB} we compute the expected sensitivity for the full data set when two operators are switched on simultaneously. In each row the left (right) panel shows to the result when the relative sign between the two operators introduced is positive (negative), since this either allows or forbids a destructive interference for the production and decay mechanisms, depending on the operators considered. For concreteness, these results are obtained assuming 20\% systematic uncertainties on both signal and background, and for a fixed mass\footnote{Data files for several values of the ALP mass can be found as ancillary material to this submission on the \texttt{arXiv} server.} of the ALP ($m_a = 200$~MeV).

The first row in Fig.~\ref{fig:cB} shows the sensitivity when the two operators that control the production of the ALP, $c_\phi$ and $c_W$, are simultaneously present. As can be seen, in the limit $c_W \to 0$ ($c_\phi \to 0)$ we recover the sensitivity to $c_\phi$ ($c_W$) given in Fig.~\ref{fig:micro-sens}, as expected. In the right panel we see that, as $c_\phi$ increases, the sensitivity to $c_W$ is enhanced because of the larger ALP production rate, see Eq.~\eqref{eq:kdij}. However, in the left panel two destructive interference patterns take place. First, when the magnitude of the two coefficients is the same (along the diagonal dotted line) the sensitivity is lost because they interfere destructively in the ALP production rate. Secondly, for $c_W \sim  m_a^2/(m_a^2-m_{\pi}^2) c_\phi \alpha / (4 s_w^2 \pi) \sim 2 c_\phi \alpha /\pi$ (along the diagonal dashed line) the contributions from the two operators to $c_{\gamma\gamma}$ cancel each other (see Eq.~\eqref{eq:cgg}), leading to a suppressed decay width into the $a\to\gamma\gamma$ decay channel. As a result, the sensitivity to the ALP using a di-photon signal is lost, but the experiment would still be sensitive to $a\to e^+ e^-$.

In the middle row of Fig.~\ref{fig:cB} we see the effect due to the interplay between $\mathcal{O}_B$ and $\mathcal{O}_\phi$. As already mentioned, since the $\mathcal{O}_B$ operator does not significantly affect the ALP production rate, it will only have an impact on its lifetime. In the right panel the two coefficients have opposite signs and, thus, no destructive interference arises in the decay width. As we can see, the inclusion of the $\mathcal{O}_B$ operator allows for a better sensitivity to $c_\phi$ up to a factor of $\sim 10$, for searches using the $\gamma\gamma$ channel. This is due to the larger decay width obtained in this case (which makes the ALP shorter-lived, increasing the decay rate within the detector) combined with the larger branching ratio into the $\gamma\gamma$ channel, see Eqs.~\eqref{eq:cll} and~\eqref{eq:cgg}. In the left panel a similar effect can be seen, leading to an increased sensitivity to $c_\phi$ for values of $c_B \sim \mathcal{O}(10^{-4})$. However since in this case the two coefficients enter with the same sign in Eq.~\eqref{eq:cgg} they can interfere destructively and suppress the ALP decay width into $\gamma\gamma$, for values of $c_B \sim m_a^2/(m_a^2-m_{\pi}^2) c_\phi \alpha / (4 c_w^2 \pi) \sim 2 c_\phi \alpha / (3 \pi)$. Moreover, once $c_B$ gets too large the sensitivity to $c_\phi$ is lost as the ALP decays too fast and the decay probability is exponentially suppressed.

Finally, the lower row in Fig.~\ref{fig:cB} shows the interplay between $c_B$ and $c_W$. In this case, the absence of the $\mathcal{O}_\phi$ operator implies that the signal to $\ell^+\ell^-$ is heavily suppressed and the sensitivity is obtained only in the $\gamma\gamma$ channel. In the left panel no interference is expected since the two coefficients enter with the same sign in $c_{\gamma\gamma}$; however, for values of $c_B \gtrsim 2\times 10^{-5}$ the ALP becomes too short-lived and the sensitivity is lost. In the right panel, on the other hand, since the two couplings enter with a different sign a destructive interference can arise in $c_{\gamma\gamma}$, leading to longer lifetimes. Thus, sensitivity to larger values of $c_W$ and $c_B$ is still possible along the line where $c_W s_w^2 \sim c_B c_w^2$, see Eq.~\eqref{eq:cgg}.

\section{Summary and conclusions}
\label{sec:conclusions}

In spite of the strong experimental evidence pointing towards the existence of BSM physics, our efforts to discover it at colliders and direct detection experiments have been unfruitful so far. While it is possible that the new physics is too heavy and lies outside of our reach at the LHC, an interesting alternative is that the new physics is light but weakly coupled to the visible sector, making it very elusive. 

Neutrino experiments, counting on very massive detectors and powerful sources, lie at the edge of the intensity frontier and are therefore well-suited to search for weakly coupled light degrees of freedom. In this work we focused on the ALP scenario, which is well-motivated from the theoretical point of view: light pseudoscalars generally arise as pseudo-Nambu-Goldstone bosons of BSM theories with spontaneous breaking of a global symmetry, and might address some of the most relevant open questions in particle physics, such as the strong CP problem or the origin of dark matter. 

For concreteness, we have considered a set of higher-dimensional effective operators coupling the ALP to the electroweak gauge bosons, which would lead to ALP production from kaon decays, via $K\to\pi a$. Working in chiral perturbation theory ($\chi$PT) allows to derive the effective coupling of ALP to mesons at low energies in all generality. While this had been done in the literature before, we have obtained the relevant expressions for the particular set of operators considered in this work. In doing so, we have also shown explicitly that previous calculations taking only into account the penguin diagrams agree with the full result obtained in $\chi$PT for the case under consideration here, as they capture the leading order contributions to this process.  
 
Next, we have recasted a recent MicroBooNE analysis~\cite{MicroBooNE:2021usw} which used data taken for the NuMI beam, searching for electron-positron pairs pointing towards the NuMI absorber. These results can be directly applied to our model, if the ALP is coupled to the SM through the $\mathcal{O}_\phi$ operator. Our results show that MicroBooNE data already sets competitive bounds on this operator (comparable to those of NA62) for ALP masses between 100 and 200 MeV (Fig.~\ref{fig:micro-bound}). For completeness, we also present our results in the plane of production branching ratio vs the ALP lifetime, without taking into account that these are typically correlated within a given model (Fig.~\ref{fig:br-ctau}). Being model-independent, this allows to easily recast our constraints to other models (including a different set of ALP couplings), or even for a different long-lived particle as long as its production and decay mechanisms are the same as considered here. For optimal values of the lifetime of the long-lived particle, current MicroBooNE data sets tight constraints on the production branching ratio, $\mathrm{BR}(K\to\pi a) < \mathcal{O}(\mathrm{few} \times 10^{-11})$.

Finally, we have also computed the sensitivity using the full NuMI dataset recorded at MicroBooNE (Fig.~\ref{fig:micro-sens}). Due to the excellent particle identification and resolution capabilities of the LArTPC technology, we have presented our sensitivities for three different searches as indicated, $a\to \mu\mu$, $a\to ee$ and $a\to\gamma\gamma$. Our results show that, depending on the level of systematic uncertainties assumed, MicroBooNE might be able to improve over current constraints for masses in the range between 100 and 250~MeV. We point out the complementarity among searches using different final state topologies, which takes full advantage of the unique LAr TPC capabilities: while searches for an excess in the $\ell^+ \ell^-$ channels are mostly sensitive to the $\mathcal{O}_\phi$ operator, searches for an excess in the di-photon channel would be sensitive to the $\mathcal{O}_W$ operator instead. Finally, we also explored the possible interference effects arising when two operators are switched on simultaneously (Fig.~\ref{fig:cB}). Here it should be noted that, while the $\mathcal{O}_B$ operator does not induce ALP production from kaon decays, MicroBooNE can be sensitive to its impact on the ALP decay rate. Additional interference effects can take place in the production vertex between the $\mathcal{O}_\phi$ and $\mathcal{O}_W$ operators.

In summary, this work stands out as a clear example of the multiple capabilities of neutrino experiments to search for new physics, not only in the neutrino sector but in other sectors as well. Needless to say, that the type of analysis performed here may be applicable to other neutrino beam experiments using near detectors. An obvious example is the case of the DUNE experiment, which will also make use of the LAr TPC technology. In this case the near detector will be placed on axis with respect to the direction of the beam, resulting in a larger background level. However, the use of a gas TPC (instead of a LAr TPC) would reduce it significantly; additionally, the possibility to move the near detectors off-axis may allow to enhance the signal sensitivity. A study of the DUNE sensitivity to this scenario cannot be done without a careful assessment of the background levels and is left for future work.

\acknowledgments{We warmly thank Belen Gavela, Luca Merlo and Olcyr Sumensari for useful discussions, and Laura Molina Bueno for pointing out to us the bounds from NA64. We also thank the anonymous referee for pointing out to us several relevant constraints in previous literature. This project has received funding/support from the European Union's Horizon 2020 research and innovation program under the Marie Skłodowska-Curie grant agreement No 860881-HIDDeN, as well as from  Grants PID2019-108892RB-I00, PID2020-113644GB-I00 and and CEX2020-001007-S, funded by MCIN/AEI/10.13039/501100011033. The authors acknowledge support from Generalitat Valenciana through the plan GenT program (CIDEGENT/2018/019) and PROMETEO/2019/083. The work of PC is supported by Grant RYC2018-024240-I funded by MCIN/AEI/10.13039/501100011033 and by ``ESF Investing in your future''. }

\appendix
\section{Integrals}
\label{app:integrals} 

We define 
\begin{align}
A(\mu, \mu' ) \equiv 1 - e^{-18 U(\mu, \mu' )} \, , \label{eq:A}\\
U (\mu, \Lambda) \equiv - \int_\Lambda^\mu \frac{d\mu'}{\mu'}\frac{y_t^2(\mu')}{32 \pi^2} \, . \label{eq:U}
\end{align}
and the following integrals, which take into account the running of the SM coupling constants with the energy scale:
\begin{eqnarray}
\label{eq:Is}
I_1(\muew, \Lambda) &\equiv & \int_{\Lambda}^{\muew}\frac{d\mu' }{\mu' } A(\muew, \mu') \frac{3 \alpha_2^2(\mu' )}{8\pi^2} 
\,, \\
I_2(\muew, \Lambda) &\equiv & \int_{\Lambda}^{\muew}\frac{d\mu' }{\mu' } \left[1-A(\muew, \mu' )\right] \frac{9\alpha_2^2(\mu' )}{16\pi^2} \, , \\
I_3(\muew, \Lambda) &\equiv & \int_{\Lambda}^{\muew}\frac{d\mu' }{\mu' } A(\muew, \mu') \frac{17 \alpha_1^2(\mu' )}{72\pi^2}    \, ,\\
I_4(\muew, \Lambda) &\equiv & \int_{\Lambda}^{\muew}\frac{d\mu' }{\mu' }  \left[1-A(\muew, \mu' )\right] \frac{17\alpha_1^2(\mu' )}{48\pi^2} \, ,\\
I_5(\muew, \Lambda) &\equiv & \int_{\Lambda}^{\muew}\frac{d\mu' }{\mu' } A(\muew, \mu')A(\mu',\Lambda) 
\left[\frac{8 \alpha_s^2(\mu' )}{27\pi^2} + \frac{\alpha_2^2(\mu' )}{16\pi^2} + \frac{17^2}{54^2}\frac{ 3\alpha_1^2(\mu' )}{4\pi^2} \right] ,  \\ 
I_6(\muew, \Lambda) &\equiv & \int_{\Lambda}^{\muew}\frac{d\mu' }{\mu' } \left[1-A(\muew, \mu' )\right] A(\mu', \Lambda) 
\left[ \frac{4 \alpha_s^2(\mu' )}{9 \pi^2} + \frac{3\alpha_2^2(\mu' )}{32\pi^2} + \frac{17^2}{54^2}\frac{ 9\alpha_1^2(\mu' )}{8\pi^2} \right]
 .  \nonumber\\ \label{eq:Is6} 
\end{eqnarray}
In Eqs.~\eqref{eq:Is}-\eqref{eq:Is6}, $\alpha_1 \equiv \alpha / c_w^2$, $\alpha_2 \equiv \alpha / s_w^2$, $\alpha_s \equiv g_s^2/(4\pi)$ correspond to the different gauge coupling strength functions. 

The running of the coupling strength functions is computed solving the RGE equations $d\alpha_i(\mu)/d\ln \mu = -\beta^{(i)}_0 \alpha_i^2/(2\pi)$, where $\beta^{(i)}_0$ are the coefficients at one-loop order and $\alpha_i \in \left\{ \alpha_1, \alpha_2, \alpha_s, \alpha_t \right\}$, with $\alpha_t \equiv y_t^2/(4\pi)$. Specifically, we take the running of $\alpha_t$ to follow that of $\alpha_s$ and therefore use $\beta^{(1)}_0 = 41/6; \beta^{(2)}_0=-19/6; \beta^{(3)}_0=-7; \beta^{(t)}_0=-7$. In doing this, the cross terms between $\alpha_t$ and $\alpha_s$ are neglected. However, we believe this is a good approximation since do find a good numerical agreement with the results in Ref.~\cite{Bauer:2020jbp} where this effect has been considered.

Finally, let us point out that using Eq.~(37) in Ref.~\cite{Bauer:2020jbp} it is straightforward to show that $U(\mu,\Lambda)$ can be approximated as
\begin{equation}
\label{eq:U-approx}
U(\mu,\Lambda) \simeq \frac{1}{64\pi^2}\frac{g_2(\mu)^2 x_t}{2}\ln \frac{\Lambda^2}{\mu^2} \, ,
\end{equation}
where we have used the relations $m_t = y_t v / \sqrt{2}, v = 2 m_W / g_2$. Thus, Eq.~\eqref{eq:A-approx} follows directly from the substitution of Eq.~\eqref{eq:U-approx} into Eq.~\eqref{eq:A}.

\section{Loop functions}
\label{app:loop}

The loop functions in Eq.~\eqref{eq:cgg} read
\begin{align}
B_0= \bigg(\sum_{{f\,=\,c,t}} N_c Q_f^2\,B_1(\tau_f)-\sum_{f\,=\,b,\ell^{-}_{\alpha}}N_c Q_f^2\,B_1(\tau_f)\bigg)
\end{align}
where 
\begin{equation}
   \begin{array}{l}
    B_1(\tau) = 1 - \tau\,f^2(\tau) \,, \\
    B_2(\tau) = 1 - (\tau-1)\,f^2(\tau) \,, 
   \end{array}
\end{equation}
with 
\begin{align}
f(\tau) = \left\{ \begin{array}{ll} 
    \arcsin\frac{1}{\sqrt{\tau}} \,; &~ \tau\ge 1 \,, \\
    \frac{\pi}{2} + \frac{i}{2} \ln\frac{1+\sqrt{1-\tau}}{1-\sqrt{1-\tau}} \,; &~ \tau<1 \,.
   \end{array} \right.
\end{align}
Here, $\tau_f\equiv 4m_f^2/m_a^2$, $Q_f$ denotes the electric charge of the fermion $f$ and $N_c^f$ is the color multiplicity ($3$ for quarks, and $1$ for leptons).

\bibliographystyle{JHEP}
\bibliography{references}

\end{document}